\documentclass[a4paper,journal]{IEEEtran}
\usepackage{amsmath,amsfonts}
\usepackage{algorithmic}
\usepackage{algorithm}
\usepackage[free-standing-units,per-mode=repeated-symbol,mode=text,reset-text-series=false,text-series-to-math=true]{siunitx}
\usepackage{array}
\usepackage[caption=false,font=normalsize,labelfont=sf,textfont=sf]{subfig}
\usepackage{printlen}
\usepackage{microtype}
\usepackage{enumitem}
\usepackage{longtable}
\usepackage{booktabs}
\usepackage{tabularx}
\usepackage{tabularray}
\usepackage{textcomp}
\usepackage{stfloats}
\usepackage{afterpage}
\usepackage{placeins}
\usepackage{url}
\usepackage{verbatim}
\usepackage{graphicx}
\usepackage[table]{xcolor}
\usepackage{cite}
\usepackage{glossaries}
\usepackage[hidelinks]{hyperref}
\hypersetup{breaklinks=true}
\usepackage[capitalise,nameinlink]{cleveref}

\DeclareUnicodeCharacter{221E}{\ensuremath{\infty}}

\DeclareSIUnit\bit{bit}
\DeclareSIUnit\ge{GE}
\DeclareSIUnit[quantity-product=]\percent{\%}

\newacronym{alu}{ALU}{arithmetic logic unit}
\newacronym{asic}{ASIC}{application-specific integrated circuit}
\newacronym{cheri}{CHERI}{Capability Hardware Enhanced RISC Instructions}
\newacronym{cheriot}{CHERIoT}{Capability Hardware Extension to RISC-V for Internet of Things}
\newacronym{csr}{CSR}{control and status register}
\newacronym{dcls}{DCLS}{dual-core lockstep}
\newacronym{ecc}{ECC}{error correcting code}
\newacronym{epmp}{ePMP}{enhanced PMP}
\newacronym{fpga}{FPGA}{field-programmable gate array}
\newacronym{fpu}{FPU}{floating-point unit}
\newacronym{isa}{ISA}{instruction set architecture}
\newacronym{kmac}{KMAC}{Keccak Message Authentication Code}
\newacronym{lsu}{LSU}{load store unit}
\newacronym{otbn}{OTBN}{OpenTitan Big Number Accelerator}
\newacronym{otp}{OTP}{one-time programmable}
\newacronym{pmp}{PMP}{physical memory protection}
\newacronym{smepmp}{Smepmp}{PMP Enhancements for memory access and execution prevention on Machine mode}
\newacronym{soc}{SoC}{system-on-chip}
\newacronym{stkz}{STKZ}{stack zeroisation engine}
\newacronym{tbre}{TBRE}{background revocation engine}
\newacronym{wb}{WB}{writeback}

\begin{document}

\title{Area Comparison of CHERIoT and PMP in Ibex}

\author{
\vspace{0.5em}
Samuel Riedel\quad
Marno van der Maas\quad
John Thomson\quad
Andreas Kurth\quad
Pirmin Vogel
\\
\vspace{1em}
{\small
lowRISC C.I.C.\\%
Cambridge, United Kingdom%
}
\\
{\small\itshape%
info@lowrisc.org
}
}

\maketitle

\begin{abstract}
  Memory safety is a critical concern for modern embedded systems, particularly in security-sensitive applications. This paper explores the area impact of adding memory safety extensions to the Ibex RISC-V core, focusing on \gls{pmp} and \gls{cheriot}. We synthesise the extended Ibex\textregistered{} cores using a commercial tool targeting the open FreePDK45 process and provide a detailed area breakdown and discussion of the results.

  The \gls{pmp} configuration we consider is one with 16 \gls{pmp} regions. We find that the extensions increase the core size by 24 thousand gate-equivalent (\si{\kilo\ge}) for \gls{pmp} and \SI{33}{\kilo\ge} for \gls{cheriot}. The increase is mainly due to the additional state required to store information about protected memory. While this increase amounts to \SI{42}{\percent} for \gls{pmp} and \SI{57}{\percent} for \gls{cheriot} in Ibex's area, its effect on the overall system is minimal. In a complete \gls{soc}, like the secure microcontroller OpenTitan Earl Grey, where the core represents only a fraction of the total area, the estimated system-wide overhead is \SI{0.6}{\percent} for \gls{pmp} and \SI{1}{\percent} for \gls{cheriot}. Given the security benefits these extensions provide, the area trade-off is justified, making Ibex a compelling choice for secure embedded applications.
\end{abstract}


\glsresetall

\section{Memory Safety in Ibex-based Processors}\label{memory-safety-in-ibex-based-processors}

Ibex is a compact, efficient, and open-source RISC-V core developed by lowRISC, specifically designed for low-power and embedded applications~\cite{ibex,gallmann2021}. It supports the RV32IMCB~\cite{riscv} instruction set and features a configurable two-stage or three-stage pipeline, making it suitable for constrained environments such as microcontrollers and security-focused processors. Ibex is best known as the core of OpenTitan, a security \gls{soc} platform equipped with a wide range of security and I/O peripherals, and the world's first commercial-grade open-source silicon root of trust.

In security domains, memory safety vulnerabilities are a major source of software security issues. Reports from both Microsoft~\cite{Msrc2019} and Google~\cite{GoogleMemorySafety} have shown that approximately \SI{70}{\percent} of security-related bug fixes in Windows and Chrome stem from memory safety errors. These issues are not limited to large-scale software systems; embedded systems are also highly susceptible to memory-safety vulnerabilities. Exploiting these vulnerabilities allows an attacker to compromise the confidentiality, integrity, and authenticity of data stored and processed by the system.

To enhance memory safety, Ibex supports two distinct architectural extensions. The Ibex core maintained by lowRISC~\cite{ibex} supports RISC-V's \gls{pmp} as well as the \gls{smepmp}, also known as \gls{epmp}. In a parallel, alternative approach, Microsoft extended Ibex with \gls{cheri} for embedded devices in their CHERIoT-Ibex~\cite{cheriot-ibex} implementation.

In this paper, we examine and compare the area impact of these two extensions on the Ibex core as well as a complete \gls{soc} like the secure microcontroller OpenTitan Earl Grey.

\subsection{Physical Memory Protection (PMP)}\label{physical-memory-protection-pmp}

\emph{\Gls{pmp}} enhances memory access control and execution prevention by defining access rights for a configurable number of memory regions. These permissions are enforced according to privilege levels, helping to enforce memory isolation and to prevent unauthorised access.

RISC-V's \gls{pmp} specification~\cite{riscvprivileged} enforces rules on all privilege modes. If at least one \gls{pmp} entry is configured, all privilege modes except for machine mode are denied access to regions that do not have a corresponding rule. This, however, does not allow \gls{pmp} regions to apply to machine mode without applying to all less privileged modes as well. For example, \gls{pmp} cannot restrict machine mode access to memory that should only be accessible in user mode.

To provide a more flexible way to secure memory regions from machine mode, RISC-V specifies the \emph{\gls{smepmp}} extension. This enhancement enables more flexible and expressive access control rules by providing mechanisms to restrict access even in machine mode based on \gls{pmp} configuration. With \gls{smepmp}, developers can specify memory regions that are accessible to user-mode or supervisor-mode code but explicitly protected from access by machine-mode software. \Gls{smepmp} helps mitigate attack vectors where attackers attempt to trick high-privileged processes into accessing or executing tampered memory from lower-privileged processes.

\Gls{pmp} and \gls{smepmp} are relatively simple to implement in hardware and set up in software, making them attractive security features. However, their fixed and coarse-grained number of regions can be a limiting factor, as RISC-V's specification permits at most 64 \gls{pmp} regions, and misconfigurations can introduce vulnerabilities. While these mechanisms are straightforward to implement in simple cases, using \gls{pmp} regions effectively in software becomes challenging when managing complex access control policies or supporting fine-grained sharing between tasks. Neither mechanism is intended to provide language-level memory safety.

\subsection{CHERI and CHERIoT}\label{cheri-and-cheriot}

\Acrfull{cheri}~\cite{cheriwebsite}, originally developed by the University of Cambridge and SRI International, takes a different approach to memory safety by extending each memory pointer into a \emph{capability}. Instead of accessing memory directly through addresses, memory operations must go through these capabilities---which not only store a memory address but also enforce strict limits on which parts of memory can be accessed and how. \Gls{cheri}'s design ensures that these capabilities cannot be modified in unsafe ways, and it includes features for securely compartmentalising different parts of a program, making software more resistant to attacks. This enables fine-grained control over memory access and significantly improves security.

Although the initial \gls{cheri} work focused on memory safety for application-class cores, embedded applications often require tailored solutions for microcontrollers. To address this, Microsoft developed \emph{\gls{cheriot}}~\cite{cheriotmicro2023,cheriotwebsite}, successfully bringing the benefits of \gls{cheri} to embedded devices without significantly increasing complexity or resource demands. On top of that, \gls{cheriot} provides the primitives to create strong compartmentalisation models without the burden of backwards compatibility with application-class software, as well as efficient temporal safety enabled by simpler core designs. This makes \gls{cheriot} a compelling choice for embedded systems that need both efficiency and strong security guarantees.

In contrast to the more coarse-grained isolation mechanisms provided by RISC-V's \gls{pmp}, which defines memory access permissions over a limited number of fixed regions, \gls{cheriot} enforces fine-grained restrictions specific to each individual memory access instead of general restrictions that apply to all memory accesses. \Gls{pmp} can provide segmentation and isolation but is limited by the number of regions it has. \Gls{cheriot} allows C and C++ code to be fully memory safe in a way that would be infeasible in a \gls{pmp} setup with limited regions. This memory safety includes deterministically guaranteeing both spatial and temporal safety. This results in a more scalable solution for memory safety and fine-grained compartmentalisation, albeit at the cost of additional metadata in memory.

\section{Area Comparison of Extensions}\label{area-comparison-of-extensions}

\subsection{Evaluation Methodology}\label{evaluation-methodology}

To evaluate the area overhead of different extensions designed to mitigate memory safety issues, we synthesised Ibex using a commercial tool targeting the open-source FreePDK45 technology~\cite{freepdk45}. In our synthesis, we included only the processor and its instruction cache logic, excluding data memories and any surrounding \gls{soc} infrastructure. We applied reasonable timing constraints to the input and output ports and used timing models from OpenRAM~\cite{openram} for all SRAM macros. However, we did not include the SRAM macro of the instruction cache in our analysis, as it is highly technology-dependent and independent of the evaluated extensions. Therefore, we report only the core area to provide a detailed view of the changes within Ibex itself. In
\cref{subse:system-impact}, we will provide insights into the area impact of a surrounding system.

We evaluated three different Ibex implementations:

\begin{itemize}[label={},leftmargin=0pt]
  \item \textbf{Ibex Baseline (RV32EMCB)}
  \begin{itemize}[label=\textbullet,leftmargin=\leftmargini]
    \item 16 32-bit registers
    \item Support for the M (multiply/divide) extension with a single-cycle multiplier
    \item Support for the B (bit manipulation) extension
    \item \Gls{wb} pipeline-stage (3-stage pipeline)
    \item Dedicated Branch Target \gls{alu}
    \item Instruction cache \gls{ecc}
    \item No \gls{pmp}, \gls{cheriot}, or \gls{dcls} support
  \end{itemize}
  \item \textbf{Ibex PMP}
  \begin{itemize}[label=\textbullet,leftmargin=\leftmargini]
    \item Same features as the Baseline
    \item Additional \gls{pmp} support for 16 regions, including the \emph{\gls{smepmp}} extension
  \end{itemize}
  \item \textbf{Ibex CHERIoT}
  \begin{itemize}[label=\textbullet,leftmargin=\leftmargini]
    \item Same features as the Baseline
    \item \emph{\Gls{cheriot}} support, which includes a \gls{cheri} execute stage, checks for all memory accesses, the \gls{tbre}, capability load filter, and expanded register file. The core does not include the optional \gls{stkz}.
  \end{itemize}
\end{itemize}

\begin{table}[tbh]
  \caption{Ibex core area and overhead with different extensions.}
  \label{tab:ibex-area}
  \centering
  \begin{tblr}{
    colspec={X[l]X[c]l},
    width=\linewidth-10pt,
    rowhead = 1,
    hline{1,Z}= {0.8pt},
    hline{2}  = {0.5pt},
    row{1}    = {bg=white, font=\bfseries},
  }
    Configuration & Area (kGE) & Area Overhead (\%)          \\
    Ibex          & 57         & Baseline                    \\
    Ibex+PMP      & 81         & Baseline +\SI{42}{\percent} \\
    Ibex+CHERIoT  & 90         & Baseline +\SI{57}{\percent} \\
  \end{tblr}
\end{table}

\subsection{Results}\label{results}

The areas, measured in kilo gate-equivalents (kGE)\footnote{Under the tooling used in this work, specifically targeting the FreePDK45 process and using the OpenRAM memory macro compiler, a \SI{4}{\kibi\byte} SRAM macro (1024 rows, 32-bit words, byte-wise write enable, 1 read/write port) corresponds to roughly \SI{63}{\kilo\ge}, which translates to a density of \SI{1.94}{\ge\per\bit}. The size of the baseline Ibex would therefore correspond to roughly \SI{3.6}{\kibi\byte} of SRAM memory, with \gls{cheriot} adding another \SI{2}{\kibi\byte} on top. This is intended as a rough rule of thumb to aid comparison. It is important to note that this translation factor is highly dependent on the technology node, the SRAM implementation, and the SRAM macro size and configuration.}, are reported in \cref{tab:ibex-area}. Note, in the FreePDK45 technology, one gate (NAND2\_X1) corresponds to \SI{0.798}{\micro\meter\squared}~\cite{freepdk45nand}. In its embedded configuration, Ibex itself is relatively small. To enable protection against memory vulnerabilities, both the \gls{pmp} and \gls{cheriot} extensions introduce an overhead roughly equivalent to half the core size. The \gls{cheriot}-enabled core is \SI{11}{\percent} larger than the \gls{pmp}-enabled core in the configurations we investigated.

\subsection{Area Breakdown}\label{area-breakdown}

\begin{figure}[htbp]
  \centering
  \includegraphics[width=\linewidth]{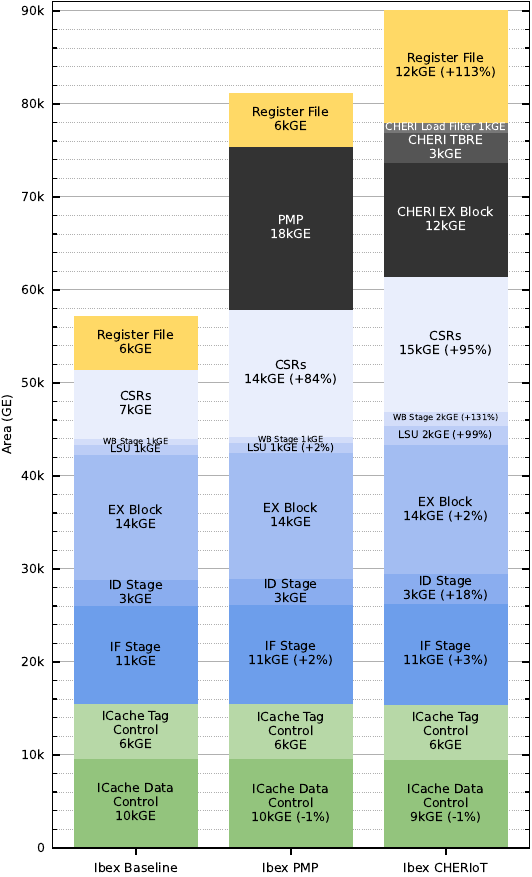}
  \caption{Area comparison between the Baseline, the \gls{pmp}-enabled, and the \gls{cheriot}-enabled Ibex cores, with the area increase annotated relative to the Baseline Ibex for modules that change more than \SI{1}{\percent}. The modules in black are modules unique to the specific configuration.}
  \label{fig:baselinevspmpvscheriot}
\end{figure}

To analyse the overhead sources in detail, we provide an area breakdown in \cref{tab:ibex-breakdown} and \cref{fig:baselinevspmpvscheriot}, comparing the Baseline, \gls{pmp}, and \gls{cheriot} extensions. In the figure, each block is annotated with the area in \si{\kilo\ge} and a percentage increase to the baseline version if the module changed more than \SI{1}{\percent}, while black boxes represent modules specific to the respective extension. The coloring of the blocks follows the hierarchy in the source code, with blue boxes making up the \emph{core} module.

We first focus on the comparison between the \gls{pmp}-enabled core and the baseline core. The \gls{pmp} core's area increase comes primarily from two modules:

\begin{table*}[tbh]
  \caption{Area breakdown of the Baseline, \gls{pmp}, and \gls{cheriot}-enabled Ibex cores. The table shows the area of each block in \si{\kilo\ge}, along with the percentage overhead compared to the Baseline configuration.}
  \label{tab:ibex-breakdown}
  \centering
  \begin{tblr}{
    colspec={X[2,l]|[0.3pt,dotted]X[r]|[0.3pt,dotted]X[r]X[r]|[0.3pt,dotted]X[r]X[r]},
    width=\textwidth-10pt,
    row{even} = {gray!20},
    rowhead = 1,
    hline{1,Z}  = {0.5pt},
    hline{3}    = {0.3pt},
    row{1,2}    = {bg=white, font=\bfseries},
  }
    \SetCell[r=2]{c} Block                                & \SetCell[c=1]{c} Ibex Baseline & \SetCell[c=2]{c} Ibex Baseline+PMP &  & \SetCell[c=2]{c} Ibex Baseline+CHERIoT & \\
                                                          & Area (kGE)                     & Area (kGE)     & Overhead (\%)        & Area (kGE)     & Overhead (\%)           \\
    \texttt{Ibex}                                         & 57.3                           & 81.4           & \SI{42.1}{\percent}  & 90.3           & \SI{57.5}{\percent}     \\
    \texttt{\textbar{}-Core}                              & 36.0                           & 60.2           & \SI{67.2}{\percent}  & 62.7           & \SI{74.2}{\percent}     \\
    \texttt{\textbar{}\ \ \textbar{}-CSRs}                & 7.4                            & 13.7           & \SI{84.2}{\percent}  & 14.5           & \SI{95.1}{\percent}     \\
    \texttt{\textbar{}\ \ \textbar{}-IF\ Stage}           & 10.5                           & 10.7           & \SI{2.1}{\percent}   & 10.8           & \SI{2.8}{\percent}      \\
    \texttt{\textbar{}\ \ \textbar{}-ID\ Stage}           & 2.8                            & 2.8            & \SI{0.4}{\percent}   & 3.3            & \SI{18.0}{\percent}     \\
    \texttt{\textbar{}\ \ \textbar{}-EX\ Block}           & 13.5                           & 13.5           & \SI{0.0}{\percent}   & 13.8           & \SI{2.2}{\percent}      \\
    \texttt{\textbar{}\ \ \textbar{}-LSU}                 & 1.1                            & 1.1            & \SI{2.1}{\percent}   & 2.1            & \SI{99.1}{\percent}     \\
    \texttt{\textbar{}\ \ \textbar{}-WB\ Stage}           & 0.7                            & 0.7            & \SI{0.4}{\percent}   & 1.5            & \SI{130.8}{\percent}    \\
    \texttt{\textbar{}\ \ \textbar{}-PMP}                 & -                              & 17.5           & ∞                    & -              & -                       \\
    \texttt{\textbar{}\ \ \textbar{}-CHERI\ EX\ Block}    & -                              & -              & -                    & 12.3           & ∞                       \\
    \texttt{\textbar{}\ \ \textbar{}-CHERI\ TBRE}         & -                              & -              & -                    & 3.2            & ∞                       \\
    \texttt{\textbar{}\ \ \textbar{}-CHERI\ Load\ Filter} & -                              & -              & -                    & 1.1            & ∞                       \\
    \texttt{\textbar{}-ICache\ Data\ Control}             & 9.6                            & 9.5            & \SI{-0.7}{\percent}  & 9.5            & \SI{-1.1}{\percent}     \\
    \texttt{\textbar{}-ICache\ Tag\ Control}              & 5.9                            & 5.9            & \SI{0.1}{\percent}   & 5.9            & \SI{0.3}{\percent}      \\
    \texttt{\textbar{}-Register\ File}                    & 5.7                            & 5.7            & \SI{-0.2}{\percent}  & 12.2           & \SI{112.5}{\percent}    \\
  \end{tblr}
\end{table*}

\begin{enumerate}
  \item The \textbf{\gls{pmp} block}, which implements \gls{pmp} checking for 16 regions in Ibex's pipeline stage, is approximately \SI{18}{\kilo\ge} in size---comparable to Ibex's \gls{alu}.
  \item The \textbf{\glspl{csr}} also contribute significantly. To store \gls{pmp} configurations and protected address ranges, multiple \glspl{csr} must be added, depending on the number of regions. Supporting 16 address regions increases the CSR size by a factor of 1.8$\times$.
\end{enumerate}

The \gls{cheriot}-enabled core exhibits a similar overhead pattern to the \gls{pmp} one:

\begin{enumerate}
  \item New instruction logic for the \gls{cheri} Execution Stage, \glspl{tbre}, and the load filter contributes \SI{16}{\kilo\ge}.
  \item \glspl{csr} increase due to additional registers tracking system state, enlarging the CSR block by a factor of 1.9$\times$.
  \item Register file size doubles to accommodate the additional capability registers.
  \item The \gls{lsu} and \gls{wb} stages grow to handle the extra functionality. While they double in size, their absolute contribution to the total overhead remains small, as these modules are inherently small.
\end{enumerate}

Finally, comparing the \gls{pmp} and \gls{cheriot}-enabled cores directly reveals that both require additional execution blocks and \glspl{csr}, resulting in similar core sizes. The key difference lies in \gls{cheriot}'s extension of the register file, which ultimately makes the \gls{cheriot} core \SI{11}{\percent} larger than the \gls{pmp} core.

\subsection{System Impact}\label{subse:system-impact}

So far, we have primarily discussed the impact of memory safety extensions on the Ibex core. However, in a complete system, the core typically occupies only a small fraction of the total chip area. Components such as the instruction and data paths, interconnects, peripherals, and accelerators quickly take up significantly more space---especially with respect to small processors like Ibex. As a result, even a doubling of core size would have only a minor impact on the overall chip area.

\subsubsection{Earl Grey Area Breakdown}\label{earl-grey-area-breakdown}

\begin{figure*}[p]
  \centering
  \includegraphics[width=0.915\linewidth]{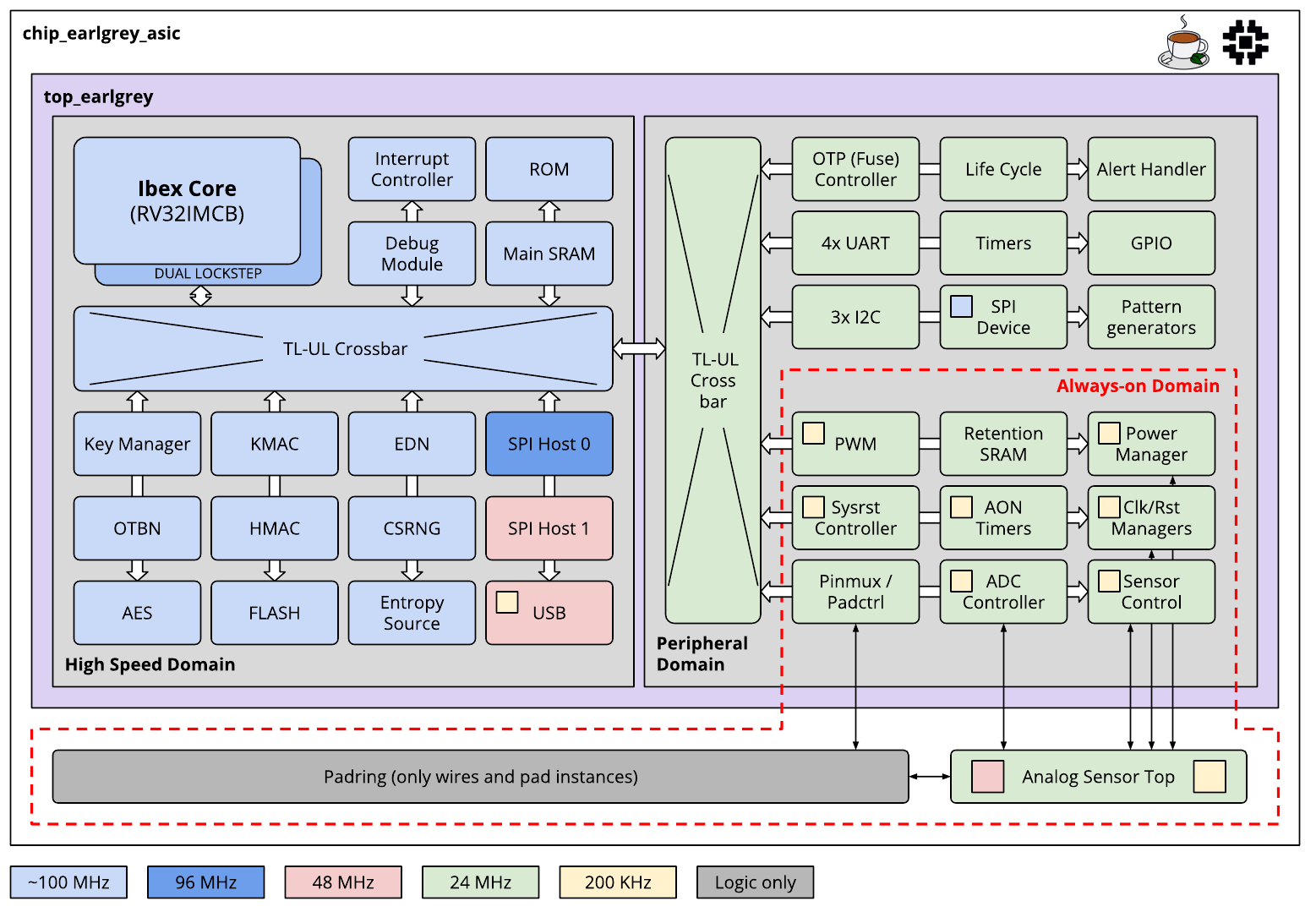}
  \caption{Block diagram of the OpenTitan Earl Grey \gls{soc}~\cite{opentitanDocs}. The synthesised top level discussed in this paper is coloured in purple.}
  \label{fig:earlgrey_block_diagram}
\end{figure*}

To assess the overall impact on the system, we synthesised the OpenTitan Earl Grey~\cite{opentitanDocs} top level, i.e., the discrete chip implementation of OpenTitan. A block diagram of Earl Grey is shown in \cref{fig:earlgrey_block_diagram}, with the synthesised top level discussed here coloured in purple.

\begin{table}[tb]
  \caption{Logic area breakdown of the OpenTitan Earl Grey based system with the baseline Ibex core presented here. The table shows the logic area of each block in \si{\kilo\ge}, along with the percentage contribution of each block.}
  \label{tab:ot-breakdown}
  \centering
  \begin{tblr}{
    colspec={X[l]rr},
    row{even} = {gray!20},
    rowhead = 1,
    hline{1,Z}  = {0.5pt},
    hline{3}    = {0.3pt},
    row{1,2}    = {bg=white, font=\bfseries},
  }
  \SetCell[r=2]{l} Block                                      & Area    & Contribution         \\
                                                              & (kGE)   & (\%)                 \\
  \texttt{Earl Grey based system}                             & 2060.75 & \SI{100.00}{\percent}\\
  \texttt{\textbar{}-Baseline Ibex Core}                      & 57.32   & \SI{2.78}{\percent}  \\
  \texttt{\textbar{}-Security Peripherals}                    & 1289.99 & \SI{62.60}{\percent} \\
  \texttt{\textbar{}\ \ \textbar{}-AES}                       & 110.29  & \SI{5.35}{\percent}  \\
  \texttt{\textbar{}\ \ \textbar{}-Alert handler}             & 77.52   & \SI{3.76}{\percent}  \\
  \texttt{\textbar{}\ \ \textbar{}-CSRNG}                     & 135.18  & \SI{6.56}{\percent}  \\
  \texttt{\textbar{}\ \ \textbar{}-EDN (x2)}                  & 47.68   & \SI{2.31}{\percent}  \\
  \texttt{\textbar{}\ \ \textbar{}-Entropy Source}            & 99.71   & \SI{4.84}{\percent}  \\
  \texttt{\textbar{}\ \ \textbar{}-HMAC}                      & 73.49   & \SI{3.57}{\percent}  \\
  \texttt{\textbar{}\ \ \textbar{}-KMAC}                      & 197.14  & \SI{9.57}{\percent}  \\
  \texttt{\textbar{}\ \ \textbar{}-Key Manager}               & 93.58   & \SI{4.54}{\percent}  \\
  \texttt{\textbar{}\ \ \textbar{}-Life Cycle Controller}     & 28.03   & \SI{1.36}{\percent}  \\
  \texttt{\textbar{}\ \ \textbar{}-OTBN}                      & 319.77  & \SI{15.52}{\percent} \\
  \texttt{\textbar{}\ \ \textbar{}-OTP Fuse Controller}       & 107.59  & \SI{5.22}{\percent}  \\
  \texttt{\textbar{}-IO Peripherals}                          & 316.20  & \SI{15.34}{\percent} \\
  \texttt{\textbar{}\ \ \textbar{}-Pinmux/Padctrl}            & 46.52   & \SI{2.26}{\percent}  \\
  \texttt{\textbar{}\ \ \textbar{}-GPIO}                      & 8.09    & \SI{0.39}{\percent}  \\
  \texttt{\textbar{}\ \ \textbar{}-I2C (x3)}                  & 45.75   & \SI{2.22}{\percent}  \\
  \texttt{\textbar{}\ \ \textbar{}-SPI (x3)}                  & 152.84  & \SI{7.42}{\percent}  \\
  \texttt{\textbar{}\ \ \textbar{}-UART (x4)}                 & 45.84   & \SI{2.22}{\percent}  \\
  \texttt{\textbar{}\ \ \textbar{}-USB Device}                & 17.16   & \SI{0.83}{\percent}  \\
  \texttt{\textbar{}-Memory Controllers}                      & 160.53  & \SI{7.79}{\percent}  \\
  \texttt{\textbar{}\ \ \textbar{}-Retention SRAM Controller} & 16.62   & \SI{0.81}{\percent}  \\
  \texttt{\textbar{}\ \ \textbar{}-Flash Controller}          & 111.25  & \SI{5.40}{\percent}  \\
  \texttt{\textbar{}\ \ \textbar{}-Main SRAM Controller}      & 16.34   & \SI{0.79}{\percent}  \\
  \texttt{\textbar{}\ \ \textbar{}-ROM Controller}            & 16.33   & \SI{0.79}{\percent}  \\
  \texttt{\textbar{}-Others}                                  & 226.03  & \SI{10.97}{\percent} \\
  \texttt{\textbar{}\ \ \textbar{}-AON Managers}              & 116.57  & \SI{5.66}{\percent}  \\
  \texttt{\textbar{}\ \ \textbar{}-Debug Module}              & 16.44   & \SI{0.80}{\percent}  \\
  \texttt{\textbar{}\ \ \textbar{}-Misc}                      & 11.48   & \SI{0.56}{\percent}  \\
  \texttt{\textbar{}\ \ \textbar{}-Interrupt Controller}      & 19.96   & \SI{0.97}{\percent}  \\
  \texttt{\textbar{}\ \ \textbar{}-XBar (x2)}                 & 61.57   & \SI{2.99}{\percent}  \\
  \end{tblr}
\end{table}

\begin{figure*}[p]
  \centering
  \includegraphics[width=\linewidth]{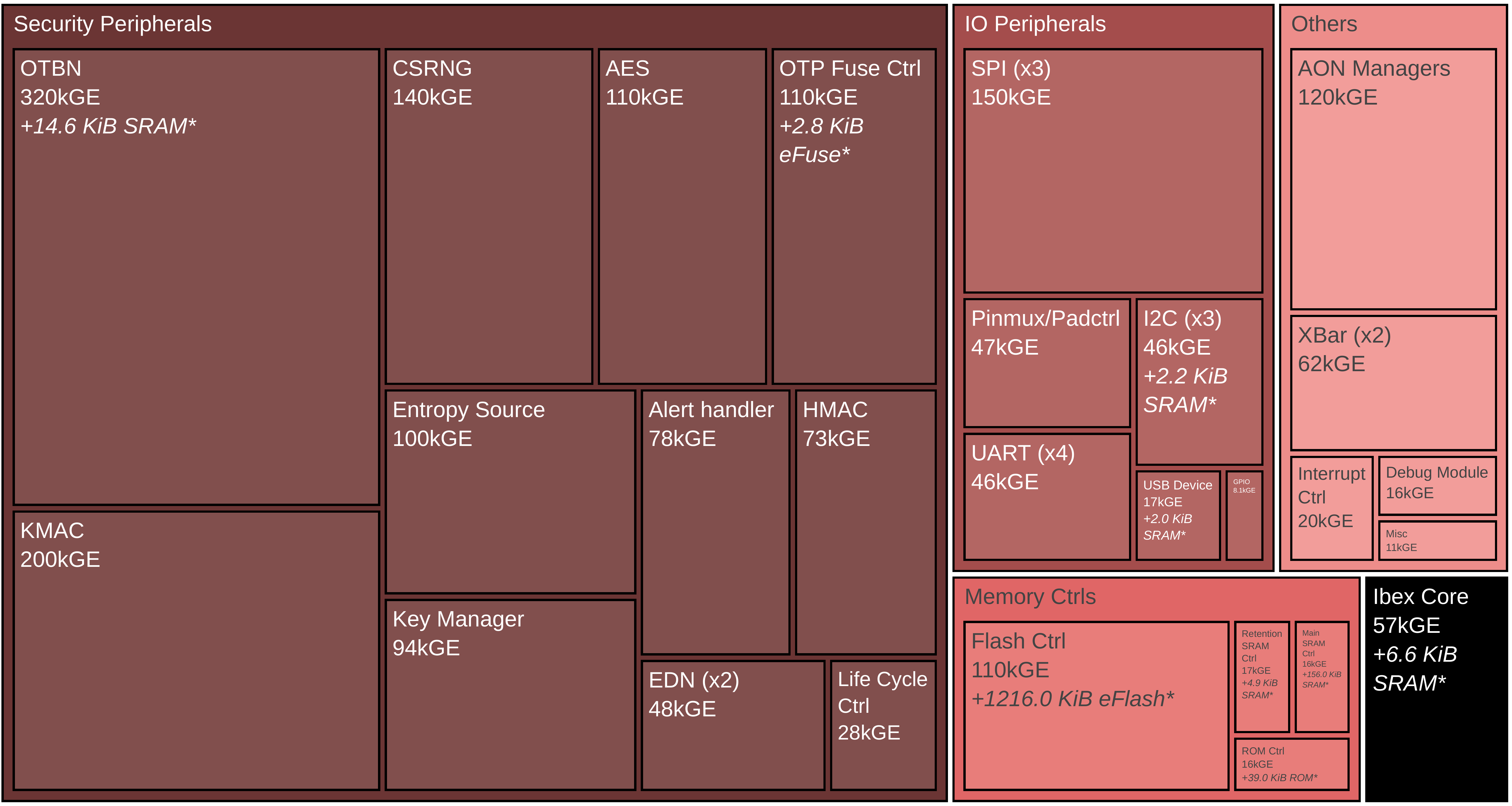}
  \raggedright{\footnotesize *Indicate the bits of memory that were \textbf{not} included in this breakdown of logic area but would be added to that module's total silicon area.}
  \caption{Area breakdown of Earl Grey Top with baseline Ibex configuration. The blocks are grouped by their functionality. The \emph{AON Managers} block comprises the PWM, Power Manager, Sysrst Controller, AON Timers, Clk/Rst Managers, ADC Controller, Sensor Control. The \emph{Misc} block summarizes the Timers and the Pattern generators.}
  \label{fig:earlgrey_breakdown}
\end{figure*}

We use the successfully taped-out v1.0.0 version of the design, which is publicly available on GitHub~\cite{opentitanGithub}, and follow the same methodology as in previous experiments, i.e., using the open-source FreePDK45 technology and targeting a main clock frequency of 100 MHz. Technology-specific and proprietary memory macros, including embedded Flash, \gls{otp} fuses, ROM, and all SRAM macros, are excluded from the logic area breakdown. As a result, the area data shown in \cref{fig:earlgrey_breakdown} and \cref{tab:ot-breakdown} only includes synthesisable logic, such as the memory controller, and not the area occupied by memory macros. For reference, the number of memory bits each module would contain is annotated in \cref{fig:earlgrey_breakdown}, providing a rough estimate of the additional area that would be required for memory. In a realistic tape-out scenario, these memory macros typically contribute an area roughly equal to that of the logic, accounting for about \SI{50}{\percent} of the total \gls{soc} area. We will discuss their impact in more detail in \cref{pmp-and-cheriot-in-earl-grey}.

The Ibex core in Earl Grey v1.0.0---featuring 32 registers, the \gls{pmp} extension with 16 ranges, and a \gls{dcls} mechanism---has a logic area of \SI{187}{\kilo\ge}, which is significantly larger than that of our baseline Ibex core at \SI{57}{\kilo\ge}. To analyse the impact of adding \gls{pmp} or \gls{cheriot} to the system, we replace the Ibex core in Earl Grey v1.0.0 with our baseline Ibex core described in \cref{area-comparison-of-extensions}. We therefore subtract the original Ibex core's logic area from Earl Grey v1.0.0's total logic area of \SI{2190}{\kilo\ge} and add our baseline Ibex in its place, resulting in a new total logic area of \SI{2061}{\kilo\ge} for the Earl Grey-based system using the baseline Ibex core. A detailed breakdown of this system's logic area is provided in \cref{tab:ot-breakdown} and \cref{fig:earlgrey_breakdown}.

The logic area breakdown reveals that a significant portion (\SI{63}{\percent}) of Earl Grey's logic area is dedicated to security peripherals, including large accelerators like the programmable \gls{otbn} and the \gls{kmac} blocks. Many of these hardware components are larger than the baseline Ibex core. The remaining logic area is distributed among I/O peripherals, memory controllers, and various supporting modules such as the interrupt controller, timers, and the debug module. Notably, the baseline Ibex core itself occupies just \SI{2.8}{\percent} of Earl Grey's total logic area.

\subsubsection{PMP and CHERIoT in Earl Grey}\label{pmp-and-cheriot-in-earl-grey}

\begin{figure}[h]
  \centering
  \includegraphics[width=\linewidth]{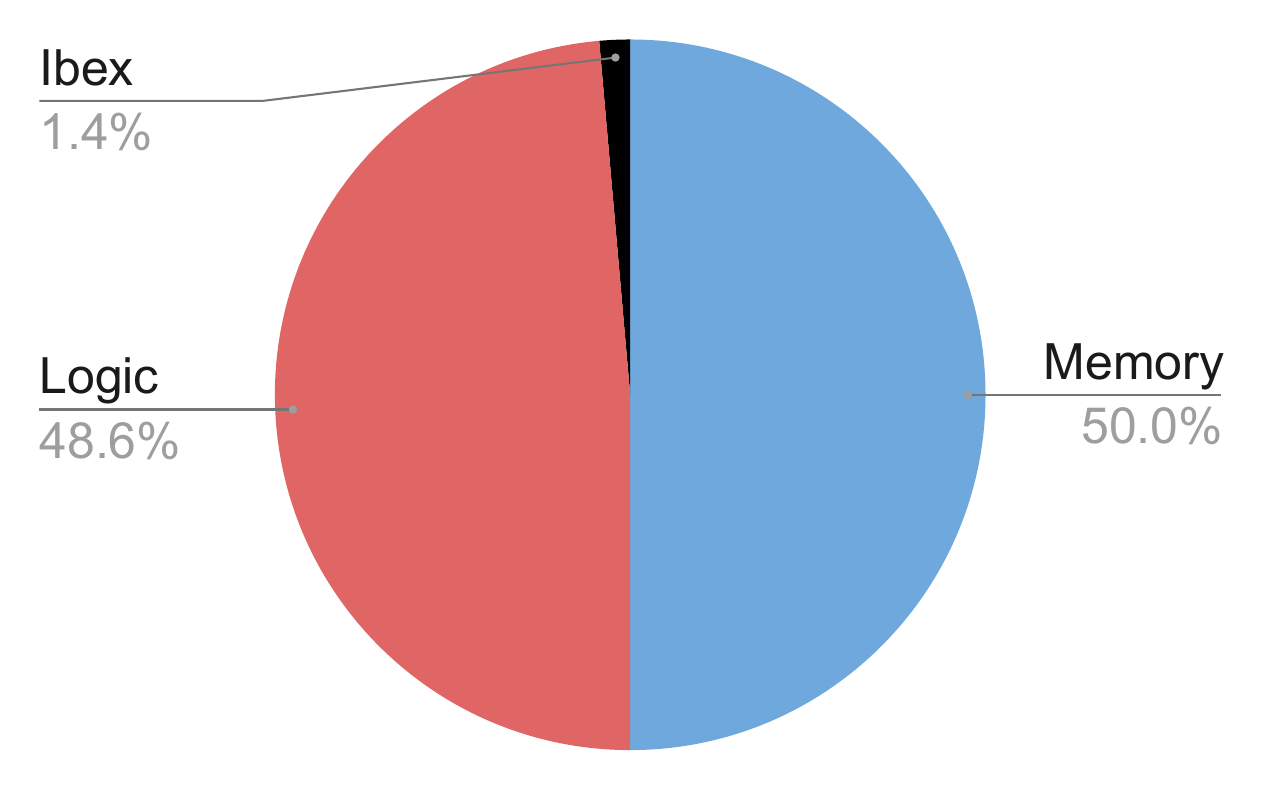}
  \caption{Earl Grey based \gls{soc} breakdown. The pie chart shows the distribution of logic and memory macros in our Earl Grey based \gls{soc}, highlighting Ibex's contribution.}
  \label{fig:soc_breakdown}
\end{figure}

When considering the full \gls{soc}, including proprietary macros, the split between logic area and these macros is approximately even, as previously noted. As a result, the contribution of baseline Ibex to the total \gls{soc} area is effectively halved compared to its share of the logic area. Specifically, while Ibex accounts for \SI{2.8}{\percent} of the logic area, this translates to just \SI{1.4}{\percent} of the overall \gls{soc} area. This breakdown is illustrated in \cref{fig:soc_breakdown}, which highlights the distribution of logic and memory macros in Earl Grey, with Ibex's contribution clearly marked.

Now that we know the area ratios of Earl Grey, we can calculate the impact of our security extensions.
Enabling the \gls{pmp} extension increases Ibex's area by \SI{42}{\percent}. Ibex then makes up \SI{3.9}{\percent} of the total logic area, and the total logic area increases by \SI{1.2}{\percent} compared to the baseline \gls{soc}. Enabling the \gls{cheriot} extension instead increases Ibex's area by \SI{58}{\percent}. Ibex then makes up \SI{4.3}{\percent} of the total logic area. In this version of the \gls{soc}, the logic area is \SI{1.6}{\percent} larger than the baseline \gls{soc} area. With respect to the total chip area, the area increase of these core extensions amounts to roughly \SI{0.6}{\percent} (\gls{pmp}) and \SI{0.8}{\percent} (\gls{cheriot}).

On top of these costs, security hardening of an \gls{soc} involves additional overhead beyond just the core. As discussed, \gls{cheriot} transforms all pointers in the system into capabilities. This requires an extra validity tag and revocation tag bit per capability, which necessitates an expansion of the data memory to accommodate these extra two bits per 64 bits of data. Furthermore, the bus connecting the core to the memory needs to be one bit wider. Specifically for Earl Grey, this means to increase the width of the main and retention SRAMs from 32 to 33 bits, which results in an overall increase in chip area of \SI{0.2}{\percent}. Together with the area increase of the core extension, the estimated cost of adding support for \gls{cheriot} in a system like OpenTitan Earl Grey is around \SI{1}{\percent}.

\section{Summary}\label{summary}

Memory safety features such as \gls{pmp} and \gls{cheriot} offer different security benefits for embedded and low-power applications. In this paper, we analyse the area overhead of incorporating them into Ibex-based processors. The results show that the \gls{cheriot} extension causes a slightly larger increase in core area compared to \gls{pmp} (\SI{57.5}{\percent} vs \SI{42.1}{\percent}), which is primarily due to the expansion of the register file. However, in a complete system such as a secure microcontroller like OpenTitan Earl Grey, the estimated impact on the overall chip area would be only \SI{0.6}{\percent} (for \gls{pmp}) and \SI{1}{\percent} (for \gls{cheriot}), as the core typically occupies a small fraction of the total system size. The adoption of these memory safety features provides enhanced protection against vulnerabilities without significantly increasing the system's area, with \gls{cheriot} additionally enforcing fine-grained spatial and temporal memory safety, as well as scalable software compartmentalisation, at only a modest area increase, making them valuable choices for systems where security is paramount.

\section{Acknowledgment}
We would like to especially thank Prof. Robert N. M. Watson from the University of Cambridge and David Chisnall from SCI Semiconductor for their valuable input and feedback on this work. We also acknowledge the broader \gls{cheri} effort by the University of Cambridge and SRI International, as well as Microsoft's implementation of \gls{cheriot} in Ibex, which made this analysis possible.

\section*{About lowRISC\textnormal{\textregistered}}
Founded in 2014 at the University of Cambridge Department of Computer Science and Technology, lowRISC is a not-for-profit company (CIC) that provides a neutral home for collaborative engineering to develop and maintain open source silicon designs and tools for the long term. The lowRISC not-for-profit structure combined with full-stack engineering capabilities in-house enables the hosting and management of high-quality projects like OpenTitan and Sunburst via the Silicon Commons\textregistered{} approach.

For more information, visit \url{https://lowrisc.org/}

\Urlmuskip=0mu plus 1mu\relax
\def\UrlBreaks{\do\/\do-}
\bibliographystyle{IEEEtran}
\bibliography{main}

\end{document}